\documentclass[preprint,12pt]{elsarticle}

\usepackage{amssymb}

\newcommand{\dphi}{\Delta\phi}
\newcommand{\pt}{p_{T}}
\journal{Annals of Physics}

\begin{document}

\begin{frontmatter}



\title{d+Au Hadron Correlation Measurements from PHENIX}


\author{Anne M. Sickles, for the PHENIX Collaboration}

\address{Physics Department, Brookhaven National Laboratory, Upton NY}

\begin{abstract}
Recent observations of extended pseudorapidity correlations at the LHC in
p+p and p+Pb collisions are of great interest.
Here we present related results from d+Au collisions
at PHENIX.  We present the observed $v_2$ and
discuss the possible origin in the geometry of the
collision region.  We also present new measurements
of the pseudorapidity dependence of the ridge
in d+Au collision.  Future plans to clarify the
role of geometry in small collision systems using
$^3$He+Au collisions are discussed.

\end{abstract}

\begin{keyword}


\end{keyword}

\end{frontmatter}
The highly asymmetric nuclear collisions d+Au and p+Pb  have been studied at ultra-relativistic energies
primarily to establishing a baseline of cold nuclear matter effects
for in heavy ion collisions.  In 2003 measurements from 
RHIC~\cite{Back:2003ns,Adler:2003ii,Adams:2003im,Arsene:2003yk}
conclusively established that the jet quenching observed at RHIC~\cite{Adler:2003qi}
was due to a final state effect in the hot nuclear matter rather than an
initial state effect.  

In heavy ion collisions at both RHIC and the LHC the properties of the matter
are understood to be described by hydrodynamics with a very small sheer viscosity
to entropy density ratio, $\eta/s$~\cite{Adare:2011tg}.  
The $\eta/s$ is constrained via measurements of Fourier coefficients of the azimuthal
distribution of particles ($v_N$ where $N$ is the order of the Fourier coefficient).
Measurements of the particle pair correlations in heavy ion collisions have been well
described by products of the same $v_N$~\cite{Aamodt:2011by,ATLAS:2012at}.  One prominent feature of these correlation
functions is the so-called {\it ridge} a long range in pseudorapidity, small $\dphi$
correlation resulting from the sum of positive $v_N$ correlations around $\dphi$~=~0. 

Surprisingly, a similar long range correlation was seen in very high multiplicity p+p
collisions at the LHC~\cite{Khachatryan:2010gv} where a hydrodynamical system was not
generally expected to be created.  
Recently, at the LHC a double ridge structure has been observed also in p+Pb collisions
at 5.02~TeV~\cite{Abelev:2012ola,Aad:2012gla}.  Once the jet and dijet correlations were subtracted
out the residual distribution was largely described by a $\cos2\Delta\phi$ modulation.
Extractions of $v_2$ resulted in values with a similar magnitude to those
in heavy ion collisions~\cite{Abelev:2012ola,Aad:2012gla}, which was suggestive of
a similar hydrodynamic origin.
However, such correlations were also expected from
the Color Glass Condensate model~\cite{Dusling:2012wy}.

Given these exciting discoveries at the LHC, it was of course natural to ask if such effects
could be seen in d+Au collisions at RHIC and what could be understood about the physical
origin of these effects from the comparison between RHIC and LHC measurements.
A large sample of d+Au collisions
at $\sqrt{s_{NN}}$~=~200~GeV was taken in 2008.  Here report on measurements by the PHENIX collaboration
using that data.

\section{Midrapidty Correlations}
\begin{figure}
\includegraphics[width=0.4\textwidth]{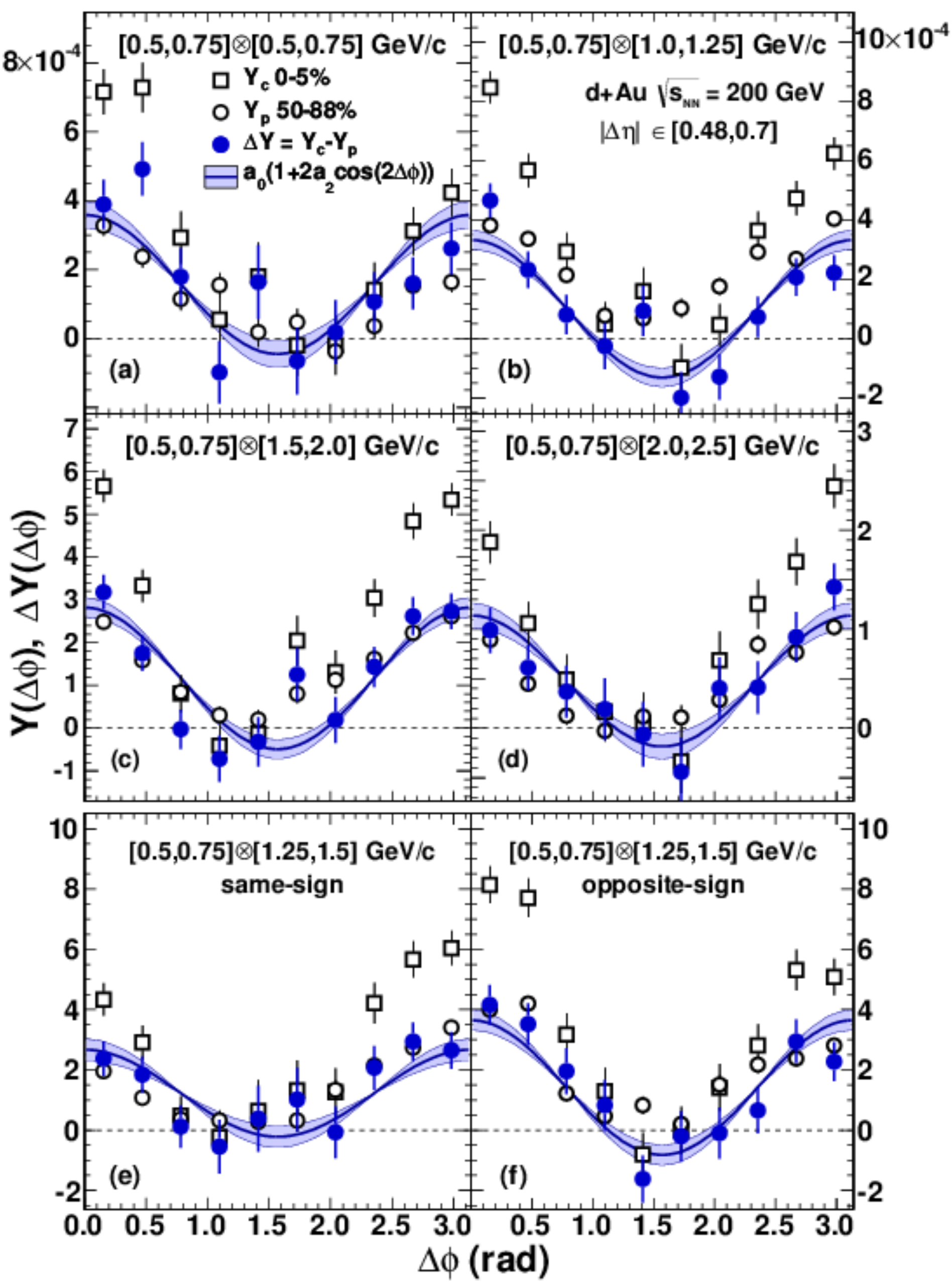}
\includegraphics[width=0.59\textwidth]{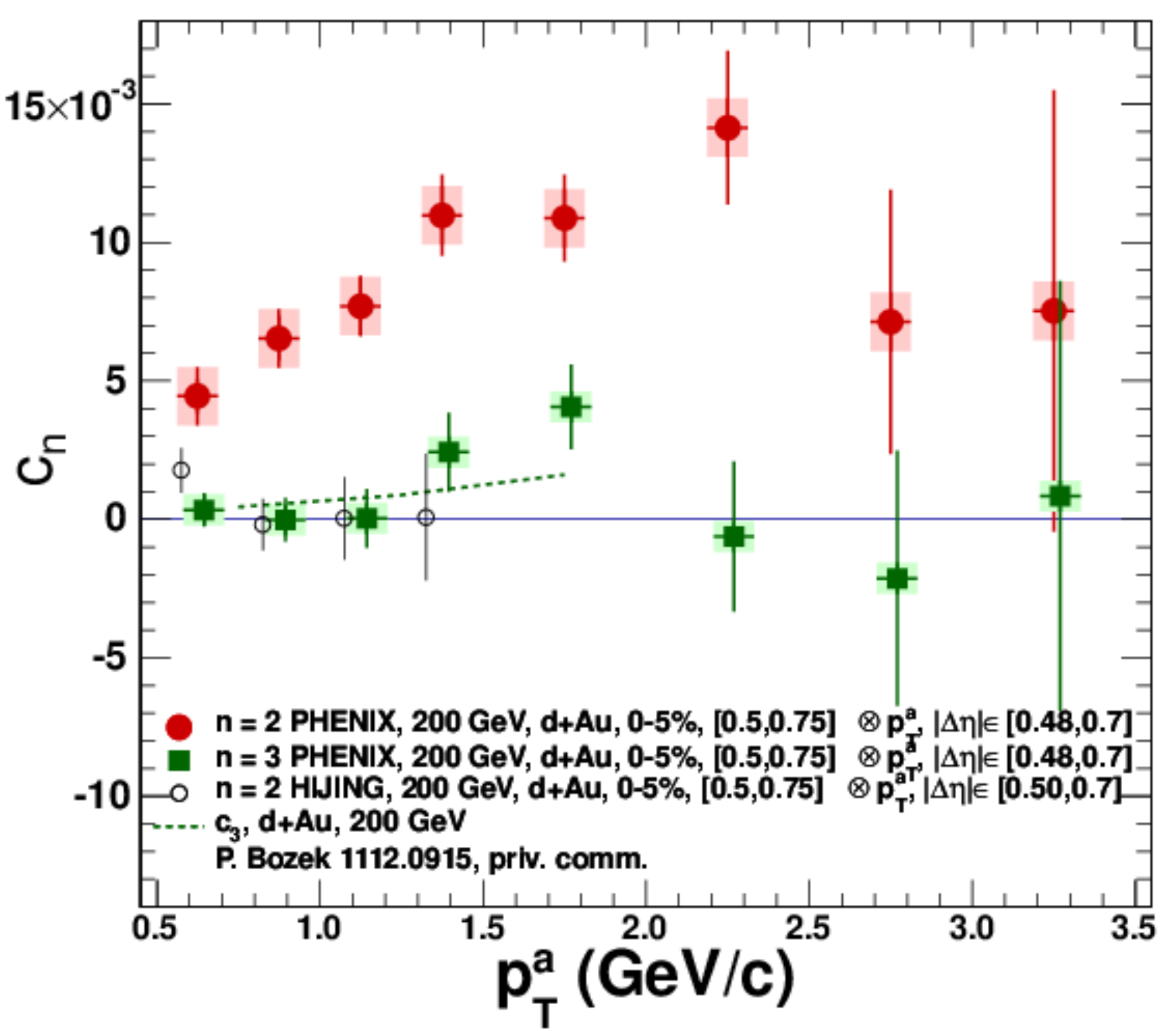}
\caption{(left) Azimuthal angular correlations between pairs of hadrons for different $\pt$
and charge combinations.  All hadrons are required to have $|\Delta\eta|>$~0.48.  Open squares
are for the 5\% most central collisions, open circles are for 50-88\% centrality and the solid 
points are the difference between central and peripheral, $\Delta Y (\Delta\phi)$.  
The underlying event has been subtracted
via the ZYAM procedure~\cite{Ajitanand:2005jj}.  Overlaid are curves showing the correlations
from the extracted $c_2$ values from the $\Delta Y ( \Delta\phi )$ distributions.
(right) $c_2$ (red) and $c_3$ (green) values extracted from the $\Delta Y \left( \Delta\phi \right)$
distributions.  The $c_2$ values from the identical procedure applied to d+Au HIJING
data is also plotted (open points) and is consistent with zero.  The $c_3$ values
are consistent with zero and with $c_3$ determined from a hydrodynamic 
calculation~\cite{Bozek:2011if,Bozek:privatecomm}. Figures are from Ref.~\cite{Adare:2013piz}.
}
\label{fig:ppg152}
\end{figure}

PHENIX has measured charged hadron azimuthal angular pair correlations in central (the top 5\% of centrality)
and peripheral (50-88\% centrality) d+Au events at $\sqrt{s_{NN}}$~=~200~GeV.  The correlations
for a selection of $\pt$ combinations are shown in the left panel of Fig.~\ref{fig:ppg152}.
The trigger particle is always 0.5$<\pt<$0.75~GeV/c and partner particle $p_T$ ($p_{T,a}$) is varied.
In order to reduce same side jet contributions a $\Delta\eta$ separation greater than 0.48 is
required between the particles.
The difference of the central and peripheral correlations, $\Delta Y (\Delta\phi)$, is also shown.
An extraction of the second Fourier component, $c_2$ is overlaid.
The curve describes the data well.  The $c_2$ values, as a function of the $\pt$ of the partner
are shown in the right panel of Fig.~\ref{fig:ppg152}.  $c_2$ reaches a maximal value of about 1\%
at around $p_{T,a}$~=~1.5~GeV/c.  Also shown on the same Figure is the third Fourier coefficient, $c_3$;
this is consistent zero in the measured $\pt$ range.  Predictions $\Delta Y$ at RHIC
in the Color Glass Condensate model have been made~\cite{Dusling:2013oia}.

\begin{figure}
\includegraphics[width=\textwidth]{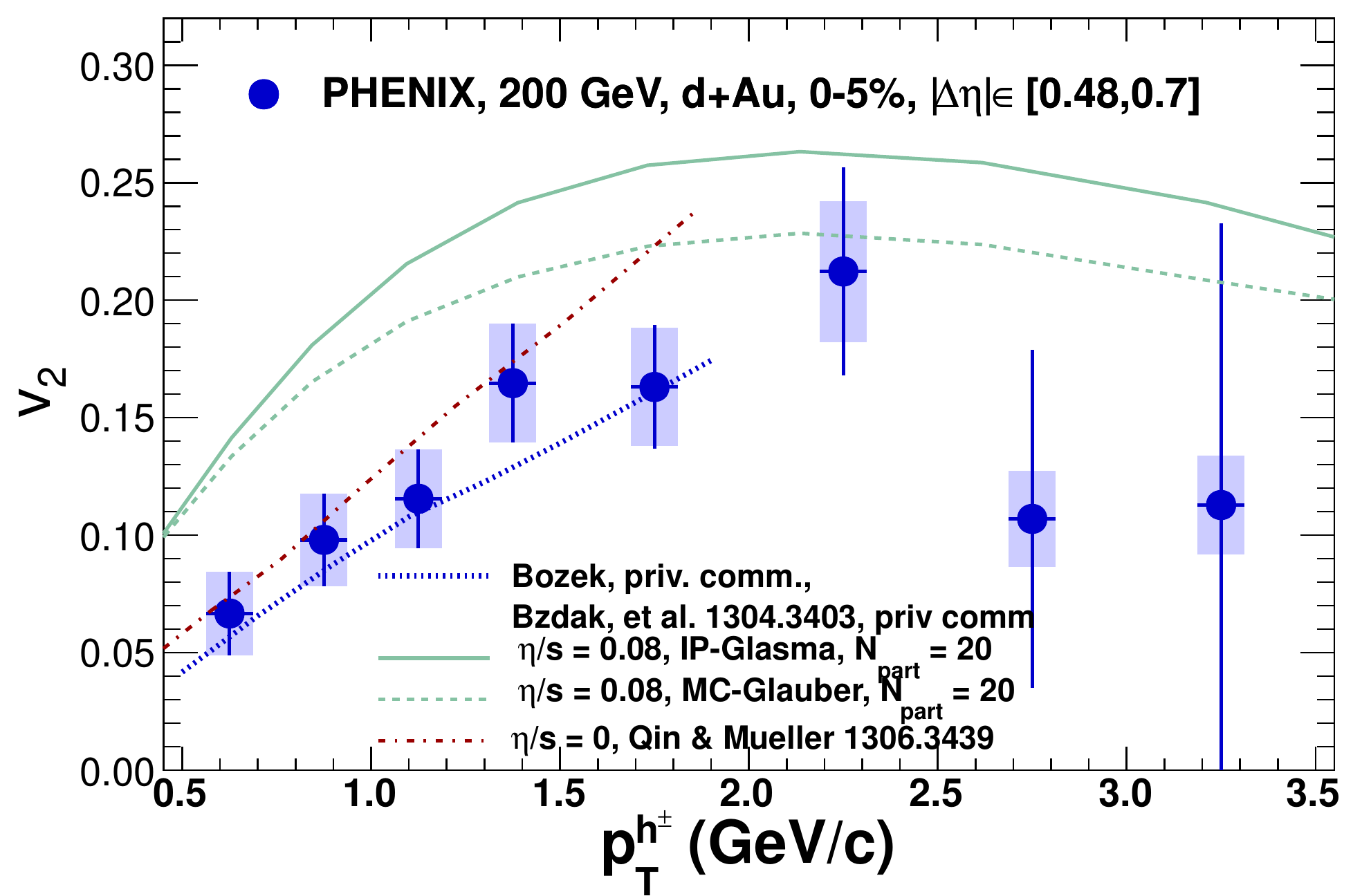}
\caption{$v_2$ as a function of $p_{T}$ for midrapidity hadrons
in the 5\% most central d+Au collisions.  Also shown on the plot are
hydrodynamic calculations from Refs.~\cite{Bozek:2011if,Bozek:privatecomm,Bzdak:2013zma,Qin:2013bha}.
Figure is from Ref.~\cite{Adare:2013piz}.}
\label{fig:v2}
\end{figure}

PHENIX lacks the pseudorapidity acceptance necessary to completely suppress same side jet correlations
(which are a small $\dphi$ and small $\Delta\eta$ effect)
in these correlations.  In order to constrain the possible effects of 
jet correlations in the observed signal
we repeat the same procedure with d+Au events generated with HIJING~\cite{Gyulassy:1994ew}.  
The results of this study are also shown in the
right panel of Fig.~\ref{fig:ppg152}.  $c_2$ in HIJING is consistent with zero and much
smaller than that observed in the data.  Additionally, we have tested the sensitivity to
 the $|\Delta\eta|$ cut 
used by varying it from the nominal 0.48 value to 0.36 and 0.60; no significant change in the 
extracted $c_2$ value is observed.
A direct measure of the large $\Delta\eta$ correlations will 
be discussed in Sec.~\ref{sec:rapsep}.

The factorization assumption~\cite{Luzum:2010sp,Alver:2010dn,Aamodt:2011by},
\begin{equation} 
c_N\left(p_{T,trig},p_{T,a}\right) = v_N\left(p_{T,trig}\right) v_N\left(p_{T,a}\right), 
\end{equation}
is used to
extract the single particle anisotropies, $v_2$, which are shown in Fig.~\ref{fig:v2}.  $v_2$ 
rises with $\pt$  reaching a maximal value of about 15\%.  
Also shown on the plot are hydrodynamic calculations from three 
groups~\cite{Bozek:2011if,Bozek:privatecomm,Bzdak:2013zma,Qin:2013bha}.  All three
calculations agree rather well with the data.
Refs.~\cite{Bozek:2011if,Bozek:privatecomm,Bzdak:2013zma} use $\eta/s$~=~0.08.
The calculation in Ref.~\cite{Qin:2013bha} is for ideal hydrodynamics.

\begin{figure}
\includegraphics[width=\textwidth]{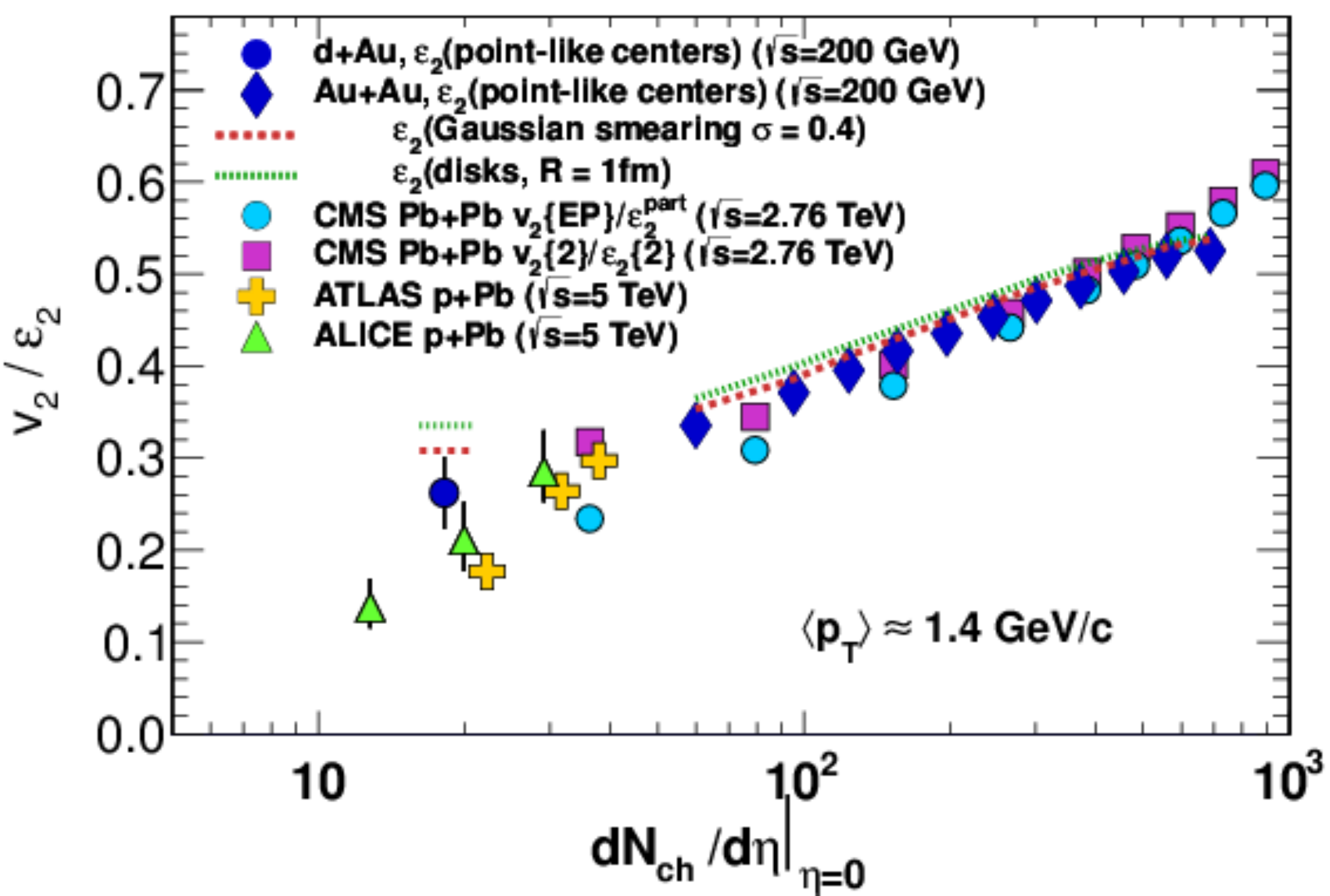}
\caption{$v_2/\varepsilon_2$ as a function of the midrapidity charged particle
multiplicity in d+Au, p+Pb, Au+Au and Pb+Pb collisions.  
The $\varepsilon_2$ values are calculated within a Glauber Monte Carlo.
Default values represent the nucleons as point-like centers. Also
shown are results with the nucleons represented as Gaussians with $\sigma$~=~0.4~fm 
(red dashed line)
and solid disks with radius of 1~fm (green dotted line).
Figure is from Ref.~\cite{Adare:2013piz}.}
\label{fig:scaling}
\end{figure}

Since the $c_3$ values observed are not significantly larger than zero, it is
not possible to extract $v_3\left(\pt\right)$ from the current data.  However,
it is possible to turn a calculation for $v_3\left(\pt\right)$ into $c_3\left(p_{T,a}\right)$, which
is shown in Fig~\ref{fig:ppg152} for the calculation in Ref.~\cite{Bozek:2011if,Bozek:privatecomm}.
The calculation agrees well with the data.

In order to investigate the possible relationship between the geometry of the collision
system and the observed $v_2$ we compare the $v_2$ values (at $\pt~\approx$~1.4~GeV/c)
scaled by estimates of the initial state second order eccentricity, $\varepsilon_2$.
The $v_2/\varepsilon_2$ values are shown as a function of $dN_{ch}/d\eta$ at mid-rapidity
in Fig.~\ref{fig:scaling} for d+Au, p+Pb, Au+Au and Pb+Pb collisions.  
The $\varepsilon_2$ values are from a Glauber Monte Carlo calculation
$v_2/\varepsilon_2$ is consistent between central d+Au and midcentral p+Pb (systems
which have a similar $dN_{ch}/d\eta$) despite the factor of 25 difference in collision energy.
The $\varepsilon_2$ value in central d+Au collisions is approximately 50\% larger than in
in midcentral p+Pb.
$v_2/\varepsilon_2$
rises as a function $dN_{ch}/d\eta$ and follows approximately a common trend between
the four collisions systems.

There are uncertainties within the Glauber Monte Carlo calculation on the $\varepsilon_2$ 
values.  We have investigated the uncertainty due to the representation of the nucleons
within the calculation.  The default values take the nucleons as point-like centers.
We have also investigated treating the nucleons as solid disks with a radius of 1~fm and
Gaussians with $\sigma$~=~0.4 in d+Au and Au+Au collisions.  In d+Au collisions
these variations change $\varepsilon_2$ by approximately 30\%.  In Au+Au the effect is much smaller.
One estimate is that there is a factor of two uncertainty on $\varepsilon_2$ in 
central p+Pb collisions~\cite{Bzdak:2013zma}.

\section{Rapidity Separated Correlations}
\label{sec:rapsep}

In order to investigate whether the correlations seen at mid-rapidity are long
range and if so, what the rapidity dependence is we have measured correlations between
mid-rapidity hadrons ($|\eta|<$0.35) and electromagnetic energy in the Muon Piston
Calorimeters~\cite{Adare:2011sc}.  These calorimeters sit on either side of the interaction region
at 3.1~$<|\eta|<$~3.7 (3.9) in the Au-going (d-going) direction.  
With this large $\Delta\eta$ gap no same side jet correlations remain.

\begin{figure}
\includegraphics[width=0.45\textwidth]{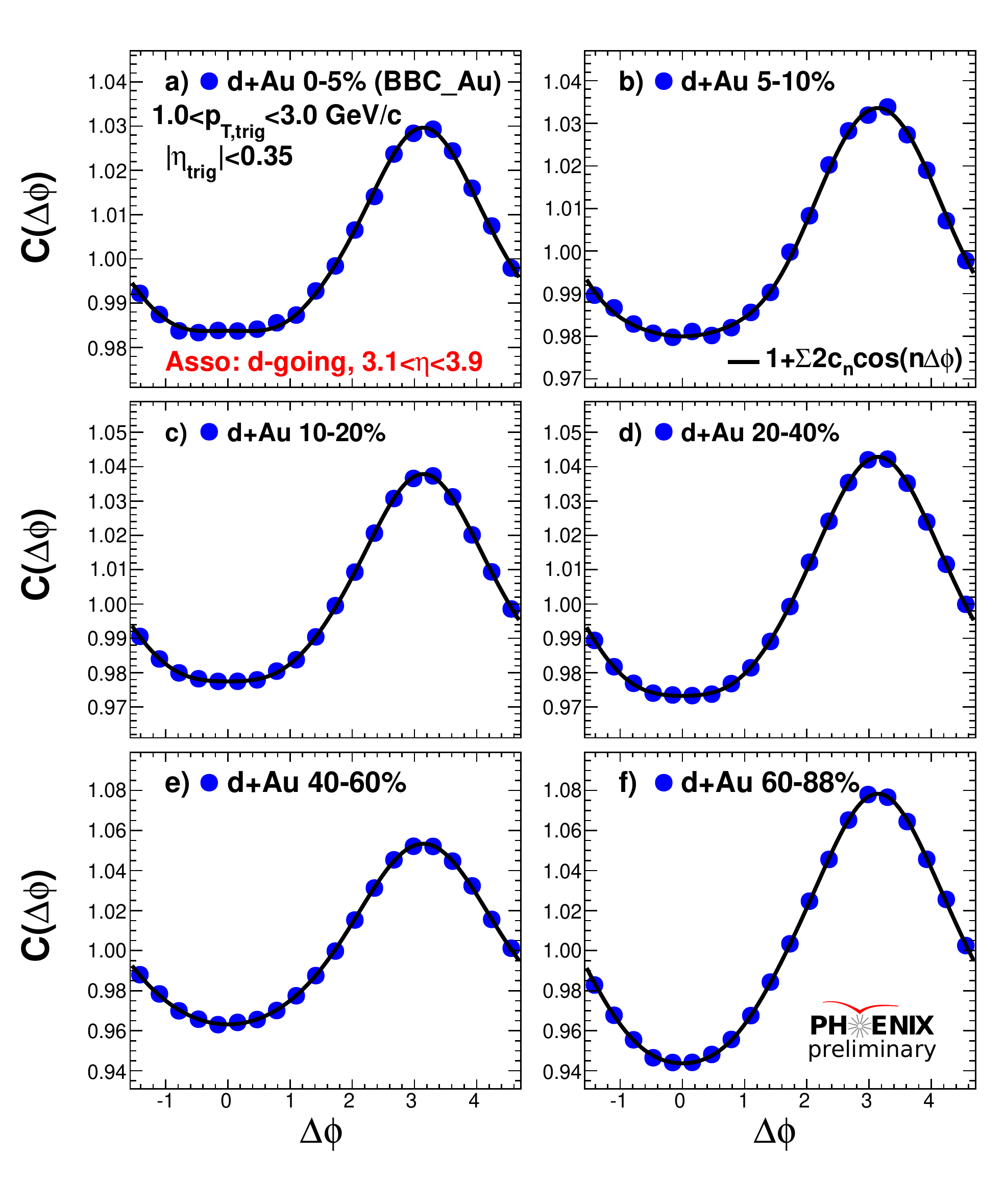}
\includegraphics[width=0.45\textwidth]{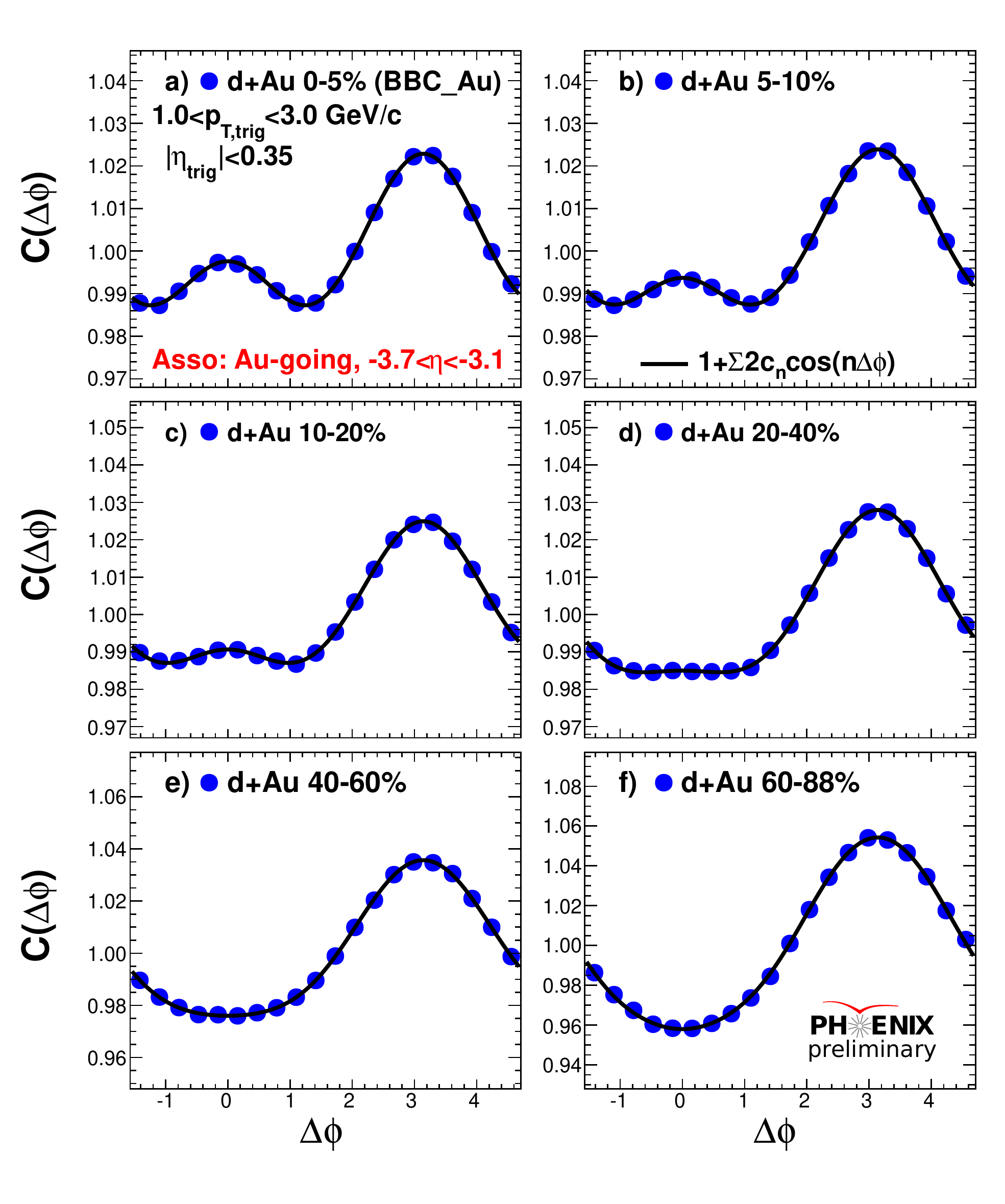}
\caption{Azimuthal correlations between mid-rapidity hadrons and $E_T$ as measured in the 
MPC in the d-going (left panels) and Au-going (right panels) direction. 
The centrality selections are as noted on the plots.}
\label{fig:mpc}
\end{figure}

The azimuthal correlations for both the deuteron going and Au-going MPC are
shown in Fig.~\ref{fig:mpc} as a function of centrality.  In all cases the dominant
feature of the correlation functions is the peak at $\dphi$=$\pi$.  This peak
has contributions from jet correlations and momentum conservation.  In the
d-going correlations no near side correlations are seen at any centrality.

In the Au-going correlations, peripheral correlations look similar to those
observed in the d-going direction.  However in central collisions a distinct
small $\dphi$ correlation is observed for the top 20\% most central events.
The magnitude of the correlation relative to the peak at $\dphi$=$\pi$
increases toward more central events. 
This is the first observation of
the ridge in d+Au collisions.
Work is ongoing to extract $c_N$ values from these correlations in the 
presence of the large $\dphi$ correlation.  

\section{Future Investigations}
\label{sec:future}

While many of the existing measurements in p+Pb and d+Au 
are suggestive of hydrodynamic behavior in d+Au collisions
further investigations of this surprising effect are certainly necessary.  As discussed above,
we see no significant $c_3$ in d+Au collisions.  In the hydrodynamic description this is consistent
with the large $c_2$ driven by the elongated shape of the deuteron and, as discussed above, the $c_3$ 
values extracted from the d+Au data are consistent with zero and much smaller than $v_2$.  If the geometry
of the deuteron drives the observed $v_2$ values then it should be possible to induce a large $c_3$
(and thereby a large $v_3$) by using a projectile nucleus with a large $\varepsilon_3$.  
PHENIX has proposed running $^3$He+Au (or $t$+Au), d+Au and p+Au collisions at RHIC in 2015
to understand the relationship between $v_N$ and geometry.

\begin{figure}
\includegraphics[width=\textwidth]{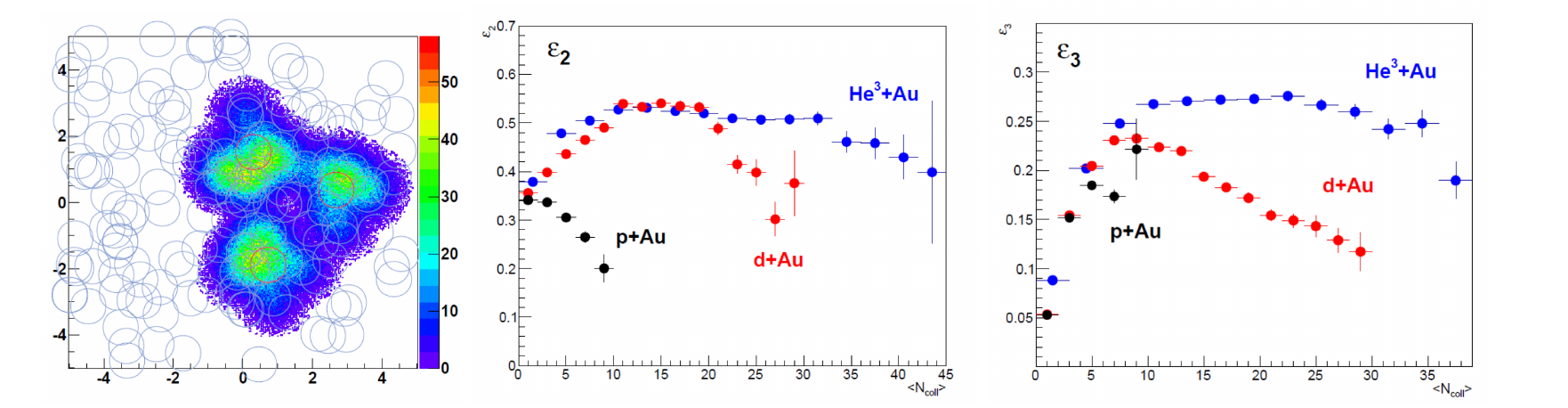}
\caption{(left)An event display of a $^3$He+Au collision as simulated in a Glauber Monte Carlo.
The nucleons are models as Gaussians with $\sigma$=0.4.  The outlines of the nucleons are shown as 
circles and the color axis shows the Glauber initial energy density. (middle) $\varepsilon_2$
as calculated in a Glauber Monte Carlo as a function of $N_{coll}$ for p+A (black), d+Au (red)
and $^3$He+Au (blue) collisions. (right) $\varepsilon_3$ as a function of $N_{coll}$.  Colors
are as in the middle panel.}
\label{fig:he3}
\end{figure}

Fig.~\ref{fig:he3} shows the results of Glauber Monte Carlo calculations of $^3$He+Au, d+Au and
p+Au collisions.  $\varepsilon_2$ decreases with an increasing number
of binary nucleon-nucleon collisions ($N_{coll}$) 
in p+Au collisions while for d+Au and $^3$He+Au collisions it increases
up to a maximal value of approximately 0.4.  For $\varepsilon_3$ $^3$He+Au collisions reaches 0.25
at about 10 collisions and remains approximately constant for all more central collisions.  
In the 0-5\% central d+Au collisions studied here the mean $N_{coll}$ value is about 18, thus
the $\varepsilon_3$ value in central $^3$He+Au collisions will be approximately 40\% bigger than
in d+Au collisions.  

Additionally, in the 2015 run PHENIX will have additional tracking provided by the silicon 
vertex detectors.  This will provide not only increase $\Delta\eta$ coverage around midrapidity, but
also increased hadron acceptance leading to more precise measurements.  
Thus, it will be possible to determine whether the increased
$\varepsilon_3$ in $^3$He+Au collisions leads to a correspondingly large $v_3$ as a conclusive test
of the role of geometry in generating the $v_N$ observed in p+A and d+A collisions.

\section{Conclusions}

Over the past year, there has been much excitement around the novel effects observed in 
p+Pb and d+Au collisions.  We have presented PHENIX results on the d+Au $v_2$ at midrapidity
and shown new results using our MPC detector which provide the first evidence for long range 
($\Delta\eta \approx$~3.5)
correlations at small $\dphi$ in d+Au collisions.  We have also discussed future plans to 
constrain the role of geometry in small collision systems by varying the shape of the 
projectile nucleus to significantly vary the $\varepsilon_3$ value of the collision region.
The collision system variation possible at RHIC make feature studies in this direction
extremely exciting at RHIC.





\bibliographystyle{elsarticle-num.bst}
\bibliography{sickles_proceedings}







\end{document}